\documentclass[a4paper,12pt]{article}
\pdfoutput=1
\usepackage[dvipdfmx]{graphicx}
\usepackage{cite}
\usepackage{amsmath}
\usepackage{amsfonts}
\usepackage{amssymb}
\usepackage{graphicx, rotating}
\usepackage{epsfig}
\usepackage{latexsym}
\usepackage{graphicx}
\usepackage{color}
\usepackage{amsmath,bm,amssymb}
\usepackage{slashed}
\usepackage{color}
\usepackage[english]{babel}
\usepackage{hyperref}

\newcommand{\Slash}[1]{{\ooalign{\hfil#1\hfil\crcr\raise.167ex\hbox{/}}}}

\newcommand{\beq}{\begin{equation}}  \newcommand{\eeq}{\end{equation}}
\newcommand{\bef}{\begin{figure}}  \newcommand{\eef}{\end{figure}}
\newcommand{\bec}{\begin{center}}  \newcommand{\eec}{\end{center}}
  
\newcommand{\laq}[1]{\label{eq:#1}}  

\newcommand{\Eq}[1]{Eq.~(\ref{eq:#1})}

\newcommand{\Sec}[1]{Sec.\ref{chap:#1}}

\def\({\left(}
\def\){\right)}

\def\O{\mathcal{O}}

\def\a{\alpha}

\def\f{\phi}

\def\l{\lambda}

\def\x{\xi}

\def\*{\dagger}

\begin{document}
\renewcommand\bibname{\Large References}

\begin{flushright}
\end{flushright}

\begin{center}

\vspace{1.5cm}

{\Large\bf {Higgs-like inflation under ACTivated mass}}
\vspace{1.5cm}

{\bf Wen Yin}

\vspace{1.5cm}

{\em Department of Physics, Tokyo Metropolitan University, Tokyo 192-0397, Japan}

\vspace{1.5cm}
\abstract{
Recent analyses that combine the latest data from the Atacama Cosmology Telescope (ACT) with cosmic microwave background observations by BICEP/Keck and Planck, together with the DESI baryon acoustic-baryonic-oscillation (BAO) measurements, have tightened the limits on inflationary scenarios. The joint data set yields a spectral index of primordial scalar perturbations  
$n_s = 0.9743 \pm 0.0034$ and an upper bound on the tensor-to-scalar ratio of $r < 0.038$.  
This slight upward shift in $n_s$ puts the previously favored Starobinsky model, and the conventional metric Higgs(-like) inflation--based on a quartic potential with a large non-minimal coupling in the Jordan frame--under tension with observations. In metric Higgs-like inflation the attractor behavior makes the predictions remarkably stable against higher-order operators, so modifying $n_s$ through such terms is difficult.  
In this paper, I show that adding a quadratic mass term to the Jordan-frame potential can raise $n_s$ and restore compatibility with the new data. 
 I also discuss how this mass term can naturally arise from threshold effects in Higgs inflation.

}
\end{center}
\clearpage
\section{Introduction}
Cosmic inflation, which generates primordial density perturbations, is strongly supported by recent observations of the cosmic microwave background (CMB). From a field-theoretical perspective, inflation is driven by a scalar field, the inflaton, whose potential must be extremely flat to sustain slow-roll dynamics. For inflation to end, however, the potential needs a slight slope so that the inflaton can roll down, decay, and reheat the Universe, thereby initiating the hot Big Bang phase of standard cosmology. The spectral tilt of the primordial scalar perturbations, imprinted in multiple cosmological data sets, directly reflects the shape of the inflaton potential, while the yet-unobserved tensor mode is expected to encode its overall height.

Recently, the latest data from the Atacama Cosmology Telescope (ACT)~\cite{ACT:2025fju,ACT:2025tim}, combined with CMB measurements by BICEP/Keck (BK)~\cite{BICEP:2021xfz} and {Planck}~\cite{Planck:2018jri,Planck:2018vyg}, together with the DESI baryon acoustic oscillation (BAO) results~\cite{DESI:2024mwx}, has shifted the preferred values to
\beq\laq{data}
n_s = 0.9743 \pm 0.0034 , \qquad r < 0.038 .
\eeq
This upward shift in $n_s$ disfavors Starobinsky inflation and conventional Higgs-like inflation, along with many other high-scale scenarios. Although several late-time cosmological tensions also hint that physics beyond the Standard Model, or new effects within the standard cosmology, with which the aforementioned preferred values may alter. However, it is good time to employ \Eq{data} to reexamine inflationary models and explore new relevant mechanisms.

One option is to adopt a low-scale inflation scenario. Axion-like inflaton models can fit the data in suitable regions of parameter space~\cite{Daido:2017wwb,Daido:2017tbr}; see also multi-natural inflation~\cite{Czerny:2014wza,Czerny:2014xja}. Low-scale hilltop inflation with a suppressed linear term is another viable possibility~\cite{Takahashi:2013cxa}. A stochastic curvaton with a non-minimal coupling that induces a Hubble-scale mass is also natural~\cite{Takahashi:2021bti}. Hybrid inflation~\cite{Copeland:1994vg,Dvali:1994ms,Linde:1997sj} can accommodate the data as well; for example, a low-scale inflaton that mixes with the QCD axion enhances the amplitude of curvature perturbations and raises the spectral index by forcing an earlier end to inflation. Natural inflation with a transient dark-energy phase can likewise match the observations~\cite{Murase:2025uwv}, and the inflaton can even act as super-heavy dark matter that explains the AMATERASU cosmic-ray event while allowing $r$ near the current upper limit and a consistent running of $n_s$.

Low-scale Higgs inflation in the Palatini formulation, where electroweak stability is maintained by a very large non-minimal coupling, also remains viable~\cite{Yin:2022fgo}. Palatini Higgs inflation is sensitive to higher-dimensional operators~\cite{Jinno:2019und,Gialamas:2019nly,Gialamas:2020vto}, so a modest shift in $n_s$ can be possible. Several recent studies have revisited this possibility in light of the ACT data~\cite{Gialamas:2025kef,Dioguardi:2025mpp,Dioguardi:2025vci}.

In this paper I focus on Higgs-like inflation in the \emph{metric} formulation~\cite{Bezrukov:2007ep,Bezrukov:2008ej}(see also~\cite{Rubio:2018ogq}). 
If the theory contains a CP-even scalar field $h$, a non-minimal coupling to gravity is generically expected,
\begin{equation}
\label{eq:nonmini}
{\cal L} \supset \frac{1}{2}\,M_{\rm pl}^2\Omega^{2} R ,
\end{equation}
where $R$ is the Ricci scalar in the Jordan frame and
\begin{equation}
\label{eq:Omega}
\Omega \equiv \sqrt{1 + \xi\,h^{2}/M_{\rm pl}^2} ,
\end{equation}
with the reduced Planck mass $M_{\rm pl}$ set to unity from now on. I assume $\xi \gg 1$ (c.f. \cite{Cook:2014dga,Hamada:2014iga}). 
Metric Higgs inflation is known to be stable against Planck-suppressed higher-dimensional operators with $\x\gg 1$~\cite{Bezrukov:2014ipa}, yet I will show that it is sensitive to a quadratic mass term in the Jordan-frame potential. 
Thus I consider
\beq
V_{\rm JF} = m^{2} h^{2} + \lambda\,h^{4}.
\eeq
In Sec.~\ref{chap:Higgsinflation}, I argue that a mass as large as $m \lesssim 1/\sqrt{\xi}$ can arise naturally through threshold effects without spoiling the unnaturally small Higgs mass after inflation. 
Before the section, I will assume that this is the inflation model, with a generic scalar field.  With the positive mass squared, strictly speaking, this is not the Higgs model, but given the similarity of the conventional Higgs inflation, $h^4$ term and non-minimal coupling, I call it Higgs-like inflation. 

Transforming to the Einstein frame yields
\beq
V = \frac{m^{2} h^{2} + \lambda\,h^{4}}{\Omega^{4}},
\eeq
and with the canonically normalized field $\phi$, defined by $(d\phi/dh)^{2} \equiv Z \equiv \Omega^{-2} + \tfrac{3}{2}\bigl(d\log\Omega^{2}/dh\bigr)^{2}$, I obtain a flat potential for $h \gg 1/\sqrt{\xi}$. As in ordinary Higgs inflation I assume the scale invariance in this regime~\cite{Bezrukov:2009db}, softly broken by $m \lesssim  1/\sqrt{\xi}$. 

In \Sec{1} I demonstrate that, unlike higher-dimensional operators, the quadratic mass term  alters the inflationary observables $n_s$ and $r$. Consequently metric Higgs-like inflation can again match current data. In \Sec{Higgsinflation}, I argue that such a mass term within the Higgs inflation can emerge from threshold corrections when $\phi \gg 1/\sqrt{\xi}$. The last section is for conclusions. 

I also comment on several approaches relevant to metric Higgs inflation in light of the ACT data. 
Reference~\cite{He:2025bli} introduces higher-dimensional kinetic terms in the Jordan frame,  
Ref.~\cite{Salvio:2025izr} adds higher-curvature operators,  
and Ref.~\cite{Aoki:2025wld} supplements the model with an extra scalar field.  
Another Higgs-like scenario with a linear non-minimal coupling is discussed in Ref.~\cite{Kallosh:2025rni}.  
In this paper, I investigate another simple and minimal possibility: I keep the usual quadratic non-minimal coupling but include an explicit mass term in the Jordan-frame potential of the Metric Higgs inflation. 

\section{Slow-roll dynamics with the mass term}
\label{chap:1}

For slow-roll inflation to happen I require both slow-roll parameters, $\varepsilon$ and $\eta$, to satisfy $\varepsilon < 1$ and $\eta < 1$.  Defining
\beq
\varepsilon(\phi) \;\equiv\; \frac{1}{2}\!
\left(\frac{\partial_\phi V}{V}\right)^{2}
\,=\, \frac{1}{2}\!
\left(\frac{\partial_h V}{V}\right)^{2}
Z^{-1},
\eeq
and
\beq
\eta(\phi) \;\equiv\; \frac{\partial^{2}_\phi V}{V}
\,=\, \frac{Z^{-1/2}\,d\!\left(Z^{-1/2}\,\partial_h V\right)\!/dh}{V},
\eeq
I see that both conditions are satisfied for sufficiently large $h$, because the potential flattens in that regime.
Indeed, expanding the Einstein-frame potential for $h \gg 1/\sqrt{\xi}$ gives
\beq
\label{eq:Vexp}
V \;\to\; \frac{\lambda}{\xi^{2}}
-\Bigl(\frac{2\lambda}{\xi^{3}}-\frac{m^{2}}{\xi^{2}}\Bigr)h^{-2}
+\Bigl(\frac{3\lambda}{\xi^{4}}-\frac{2m^{2}}{\xi^{3}}\Bigr)h^{-4}
+\cdots .
\eeq

The quadratic term affects the slow-roll dynamics as soon as $|m^{2}| \gtrsim \lambda/\xi$; hence it does \emph{not} decouple at large $\xi$ or large $h$, and I must include its effect when computing inflationary observables.

The slow-roll parameters determine the CMB observables.   
The CMB normalization of the scalar power spectrum is
$
\Delta_{\cal R}^{2}(k_{*}) \;\simeq\; 2\times10^{-9},
$
and, in the slow-roll approximation,
\beq
\Delta_{\cal R}^{2}(k_{*})
= \frac{1}{24\pi^{2}}\frac{V}{\varepsilon_{*}},
\qquad
n_{s} \simeq 1 - 6\varepsilon_{*} + 2\eta_{*},
\qquad
r \simeq 16\varepsilon_{*},
\eeq
where the subscript $*$ denotes evaluation at horizon exit for the pivot scale $k_{*}=0.05\,\mathrm{Mpc}^{-1}$.

To express the observables in terms of the e-folding number $N$, I solve the slow-roll equation $3H\dot\phi \simeq -\partial_\phi V$.  Without loss of generality I consider, $\f>0,h>0$ for the inflation. 
This gives
\beq
N \;\simeq\; \int_{\phi_{\rm end}}^{\phi_*} \!\Bigl(\frac{\partial_\phi V}{3H^{2}}\Bigr)^{-1} d\phi
\;\simeq\; \int_{h_{\rm end}}^{h_*} \!\Bigl(\frac{V}{\partial_h V}\Bigr)\,Z\,dh ,
\eeq
where $H \simeq \sqrt{V/3}$ is the Hubble parameter and $\dot X \equiv dX/dt$ with $t$ the cosmic time.  The quantities $\phi_{\rm end}$ (or $h_{\rm end}$) denote the field value at which one of the slow-roll conditions is violated.  To reduce numerical cost I impose
$
\sqrt{\varepsilon(\phi_{\rm end})^{2}+\eta(\phi_{\rm end})^{2}} = 1/3.
$\footnote{Varying this threshold between $0.3$ and $1$ changes $n_s$ by at most ${\cal O}(0.1\%)$.  I also note that the choice $\sqrt{\varepsilon^{2}+\eta^{2}}\approx1$ corresponds to $h\sim1/\sqrt{\xi}$ for relatively large $m$, that is, near the regime where the model becomes non-perturbative.  This numerical check indicates that possible non-perturbative corrections, such as the threshold effects discussed in the next section, modify the estimated $N$ by at most ${\cal O}(0.1\%)$.}

When the either two latter term in Eq.~\eqref{eq:Vexp} dominates, the well-known formula are obtained
\beq\label{eq:pred}
\Delta_{\cal R}^{2} \;\simeq\; \frac{\lambda}{18\pi^{2}}\,\frac{N^{2}}{\xi^{2}},
\quad
n_s \;\simeq\; 1 - \frac{2}{N},
\quad
r \;\simeq\; \frac{\a}{N^{2}}.
\eeq
If the ${\cal O}(h^{-2})  (\O(h^{-4}))$ term dominates, $\a=12(3)$.

In my setup not only either the ${\cal O}(h^{-2})$ or ${\cal O}(h^{-4})$ piece can dominate, because inflation occurs for $h\lesssim 10/\sqrt{\xi}$, but also both can dominate the inflation dynamics. A mild cancellation can render the ${\cal O}(h^{-4})$ term important when
$
|\,\delta| \;\lesssim\; 1,
$
with 
\beq
\delta \equiv 1 -\frac{ m^{2}\xi}{2\lambda} .
\eeq
Then
\beq
V \;\simeq\; \frac{\lambda}{\xi^{2}}
-2\l \delta\,\xi^{-3} h^{-2}
+\bigl(4\delta-1\bigr)\l\,\xi^{-4} h^{-4}.
\eeq
If $\delta=0$ (a fine tuning) or $1$, the prediction for $n_s$ reverts to Eq.~\eqref{eq:pred}.  For a mild tuning $\delta\sim 0$, the ${\cal O}(h^{-4})$ term accelerates the roll relative to the $m=0$ limit, effectively increasing $\phi_*$ for fixed $N$.  Because the curvature is smaller at larger $\phi_*$, $|\eta_*|$ decreases and $n_s$ increases.

\begin{figure}[t!]
\begin{center}
\includegraphics[width=100mm]{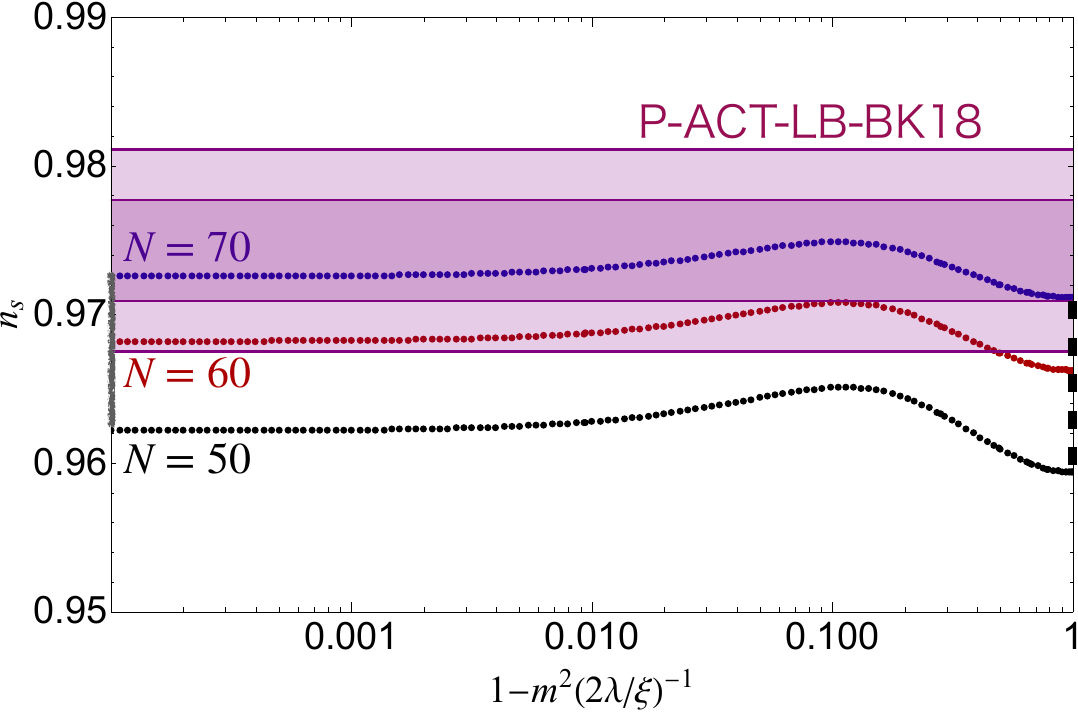}\\[5mm]
\includegraphics[width=105mm]{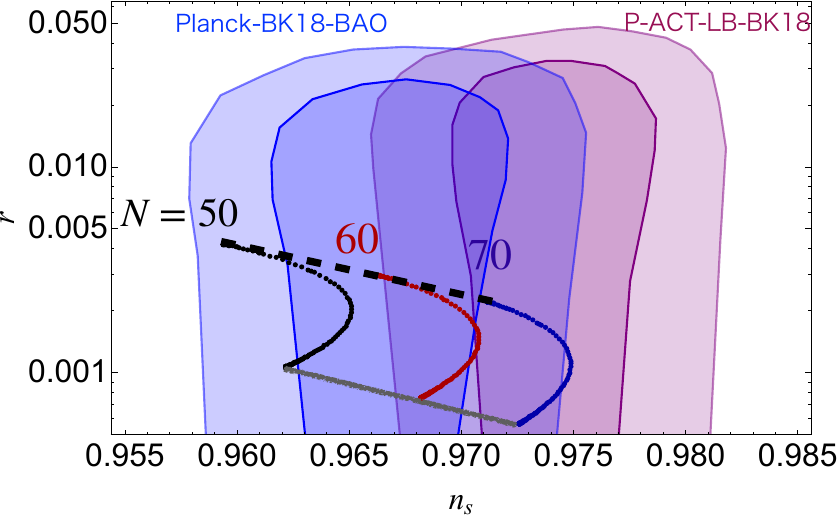}
\end{center}
\caption{Upper panel: spectral index $n_s$ versus the tuning parameter $\delta$ for $N = 50,\,60,\,70$.  The purple bands are the ACT\,+\,BK18\,+\,Planck\,+\,DESI\,BAO $1$ and $2\sigma$ regions.  Lower panel: corresponding trajectories in the $n_s$-$r$ plane with the same color coding.}
\label{fig:1}
\end{figure}

To check this behavior I solve the full equation of motion,
$\ddot\phi + 3H\dot\phi + \partial_\phi V = 0$,
and evaluate the slow-roll parameters to third order (see, e.g.,~\cite{Daido:2017tbr}).  The results are displayed in Fig.~\ref{fig:1}.\footnote{In the numerical study I fix $\xi = 10^{3}$.  Varying $\xi$ by orders of magnitudes leaves the qualitative behavior unchanged as long as $\xi\gg1$.}

The upper panel shows $n_s$ as a function of $\delta$; the three curves correspond to $N = 50,\;60,\;70$ from bottom to top.  The ACT\,+\,BK18\,+\,Planck\,+\,DESI\,BAO bounds are overlaid using the data set of Ref.~\cite{ACT:2025tim}.  The horizontal axis also measures the degree of tuning.  When $\delta=1$ ($m^{2}=0$) the model reduces to the usual $h^{4}$ Higgs inflation or Starobinsky inflation.  In the opposite limit $\delta\to0$ the model approaches the alternative regime described after Eq.~\eqref{eq:pred}.  A mild $10\%$ tuning lifts $n_s$ into the allowed region even for $N\gtrsim55$, and provides an excellent fit for $N\gtrsim60$, which can be realized with a short kination epoch. 

The lower panel shows the corresponding predictions in the $(n_s,r)$ plane for $N = 50,\,60,\,70$ (left to right).  Observational contour from combined ACT+BK18+ Planck+ DESI\,BAO, along with the combined Planck\,+\,BK18\,+\,BAO limits, are taken from Ref.~\cite{ACT:2025tim,BICEP:2021xfz}.  The dashed black line represents the standard Higgs inflation (Starobinsky) prediction, while the gray line shows the inflation by pure ${\cal O}(h^{-4})$ case.

I also checked that the running of $n_s$ and the running of running of $n_s$, respectively, are $\O(10^{-4})$ and $\O(10^{-5})$ which are negative. 

 Before closing this section I estimate the size of the mass parameter.  
Using the CMB normalization together with the mild-tuning condition $\delta\!\sim\!0$ I find
\beq
m \;\sim\; 6\pi \,
\frac{\sqrt{\Delta_{\cal R}^{2}}\sqrt{\xi}}{N}
\;\approx\; 0.0045\,
\sqrt{\frac{\Delta_{\cal R}^{2}}{2.1\times10^{-9}}}\,
\sqrt{\frac{\xi}{10^{5}}}\,
\frac{60}{N}.
\eeq
Hence $m \lesssim 1/\sqrt{\xi}$ when $\xi \lesssim 10^{5}$ (corresponding to $\lambda \lesssim 1$),  
while $m \gtrsim 1/\xi$ when $\xi \gtrsim 10^{3}$ (i.e.\ $\lambda \gtrsim 0.03$). Most of parameter region are consistent with the soft breaking assumption imposed previously. With a relatively large $\x$, a not-too-small quartic coupling can lead to the successful slow-roll inflation.

\section{Connecting the Higgs inflation}
\label{chap:Higgsinflation}

One pragmatic viewpoint is to regard the model as unrelated to the Standard Model Higgs.  
If $h$ is a new scalar, there is no reason to forbid an explicit mass term, so the Jordan-frame potential is completely natural with $m=1/\x-1/\sqrt{\x}$, which are known to be the typical energy scale of the setup~\cite{Burgess:2009ea,Barbon:2009ya,Burgess:2010zq,Bezrukov:2010jz}.

From a minimalistic perspective, however, I may instead \emph{identify} $h$ with the Standard Model Higgs field when $h\ll1/\xi$.  
The naturally large Jordan-frame mass is apparently inconsistent with the observed Higgs mass at the electroweak scale, reviving the familiar hierarchy (fine-tuning) problem.

If I insist on genuine Higgs inflation, I must also confront another difficulty: within the Standard Model the quartic coupling $\lambda$ turns negative above ${\cal O}(10^{9\text{-}12})\,$GeV (see, e.g.,~\cite{Li:2022ugn}), so a naive extrapolation of the Higgs potential leads to an anti--de Sitter region at large $h$.  
A well-known way to avoid this is to invoke threshold effects that raise the effective quartic coupling at large $h$ during inflation, rendering Higgs inflation viable even with a metastable electroweak vacuum~\cite{Bezrukov:2014ipa}.

Such threshold effects naturally arise from ultraviolet completions that restore unitarity with $h$ between $1/\xi$ and $1/\sqrt{\xi}$, where the model becomes strongly coupled~\cite{Burgess:2009ea,Barbon:2009ya,Burgess:2010zq,Bezrukov:2010jz}.  
Because the Standard Model Yukawa and Higgs couplings above that scale need not equal their low-energy values, matching across the threshold can shift the effective parameters during inflation~\cite{Bezrukov:2014ipa,George:2015nza,Fumagalli:2016lls,Enckell:2016xse,Bezrukov:2017dyv,Shaposhnikov:2020fdv,Poisson:2023tja,Wada:2024txn}.

To illustrate, consider the one-loop Coleman-Weinberg contribution in the Einstein frame\cite{Bezrukov:2014ipa}
$
\delta V(\phi) \;=\; \frac{9}{64\pi^{2}}\,
\frac{2}{\bar\epsilon}\,
\lambda^{2}\,
F^{4}\!
\left((\partial_\f F)^{2} + \frac{1}{3}F \partial^2_{\f}F\right)^{2},
$
which arises from the vacuum diagram
$\tfrac{1}{2}\,\mathrm{Tr}\,\ln\!\bigl(\Box - \partial_{\phi}^{2}(\tfrac{\lambda}{4}F^{4})\bigr)$,  
with $F\equiv h(\phi)/\Omega(\phi)$.%
\footnote{Using a momentum cutoff instead of dimensional regularization would give
$\delta V \sim \Lambda^{2}\lambda\,F^{2}((\partial_\f F)^{2} + \frac{1}{3}F \partial^2_{\f}F)$,
already hinting at a mass correction.}
The divergence demands a counterterm of the same functional form,  
$
\delta\lambda \;=\; c\,
\left((\partial_\f F)^{2} + \frac{1}{3}F \partial^2_{\f}F\right)^{2},
$
which approaches a constant $c$ for $h\ll1/\xi$ and vanishes for $h\gg1/\sqrt{\xi}$.  
Thus the inflationary plateau is almost unaffected with the high-energy quartic coupling lifted be the positive value due to negatively large enough $c$.

I point out that a similar argument applies to the Higgs mass and the cosmological constant if I write
$V_{\rm JF} = \Lambda_{cc}^{4} - m_{h}^{2}h^{2} + \lambda\,h^{4}$.  
The vacuum loop then gives
\beq
\delta V = \frac{9}{1024\pi^{2}}\,
\frac{2}{\bar\epsilon}\,
\bigl(A\,\Lambda_{cc}^{4} - m_{h}^{2}B
      + 12\lambda\,F^{2}((\partial_\f F)^{2}+F \partial^2_\f F/3)\bigr)^{2},
\eeq
with
$
A = \frac{20(\partial_\phi\Omega)^{2}}{\Omega^{6}}
    - \frac{4\,\partial_\phi^{2}\Omega}{\Omega^{5}},
\quad
B = -\frac{8\,F\,\partial_\phi F\,\partial_\phi\Omega}{\Omega^{3}}
    + \frac{2\,F\,\partial_\phi^{2}F + 2(\partial_\phi F)^{2}}{\Omega^{2}}
    + F^{2}\!\left(\frac{6(\partial_\phi\Omega)^{2}}{\Omega^{4}}
    - \frac{2\,\partial_\phi^{2}\Omega}{\Omega^{3}}\right).
$
Renormalization therefore requires independent counterterms for
$m_{h}^{2}$, $\Lambda_{cc}^{4}$, and $\lambda$,  
and the threshold corrections can shift the effective Higgs mass squared positively large during inflation what my inflationary fit demands.

At small field values $h \ll 1/\xi$, the counterterm shifts the potential by a constant
$c_{3} + c_{2}\,h^{2} + c\,h^{4}$.\footnote{The counterterm behaves as
$\propto(-m_{h}^{2}+6h^{2})^{2}/\bar\epsilon$, which resembles the original
potential once the vacuum energy is set to zero, but the degeneracy is easily
lifted at higher loops.}
In the opposite limit $h\gg 1/\sqrt{\xi}$, I find
$\delta V \propto 1/(\xi^{4}h^{4})$. 
This does not contribute to the cancellation of $h^{-2}$ term. Although it may slightly change the prediction on $m$, the prediction in $n_s, r$ plane does not change much. 
Because the strong dynamics enters between $1/\xi$ and $1/\sqrt{\xi}$, it is
natural for $c_{2}$ to be of that order.  Matching to the low-energy Higgs mass
then requires
$
c_{2}+m^{2} = -m_{h}^{2},
$ 
namely a fine tuning.\footnote{A similar tuning is needed for the cosmological
constant.  More generally, for inflation one could write
$V_{\rm JF}= \Lambda^{4}+m^{2}h^{2}+\lambda h^{4}$.  The fit to $n_s$ improves
when $\Lambda^{4}\approx m^{4}/(4\lambda)$, but in that case inflation ends at
$h\ll1/\sqrt{\xi}$, where non-perturbative effects are significant, and we need a further tuning for enhancing the spectral index, 
so I do not explore that
possibility here. 
 I have checked that as long as
$\Lambda^{4}\lesssim m^{4}/(4\lambda)$, the predictions remain unchanged.
For $m^{2}=-\sqrt{\Lambda^{4}/(4\lambda)}$, studied in
Ref.~\cite{Linde:2011nh}, I reproduce the known result that the predictions
approach the original Higgs inflation for $\xi\gg1$. This is because the
${\cal O}(h^{-2})$ term in Eq.~\eqref{eq:Vexp} cannot be suppressed when
$m^{2}<0$, while the cosmological constant only affects the
${\cal O}(h^{-4})$ piece.}

Other massive BSM states can also shift the Higgs mass threshold.
For example, a sterile neutrino with Majorana mass $M$ contributes the term
$(M/\Omega)\,N^{c}N$ in the Einstein frame.  The leading two-loop counterterm
for the Higgs mass is then proportional to 
$(\partial_\phi\Omega^{-1})^{2},$ and $ \Omega^{-1}\partial_\phi^{2}\Omega^{-1}$,
which vanish for $h\ll1/\xi$ but approaches a constant for
$h\gg1/\sqrt{\xi}$.

In summary, threshold effects readily generate a large effective Higgs mass at
high field values while leaving the low-energy Higgs mass unnaturally small, a
manifestation of the usual electroweak fine-tuning problem.  Whether this
hierarchy is resolved by the ultraviolet completion, by dynamical mechanisms,
or by anthropic selection lies beyond the scope of the present work.

\section{Conclusions}

With a large non-minimal coupling, metric Higgs inflation is famously stable
against Planck-suppressed higher-dimensional operators, so its predictions have
long been regarded as robust.  Recent data from ACT combined with BK, Planck, DESI-BAO data, however, place
those predictions in mild tension with observation.  In this paper I have
demonstrated that introducing a simple lower-dimensional operator—a quadratic
mass term in the Jordan-frame potential—can shift the spectral index enough to restore agreement with the latest data, thereby reviving metric Higgs-like inflation.

I have also argued that a mass of the required size arises naturally once
threshold effects are included in the Higgs sector: the same ultraviolet
physics that stabilizes the potential at large field values can generate a Jordan-frame mass without disturbing the low-energy Higgs mass more than the usual electroweak fine tuning. In this way the model remains minimal yet compatible with current cosmological constraints, and it
provides a concrete target for future measurement of the spectral index.

\section*{Acknowledgement}
This work is supported by JSPS KAKENHI Grant Nos. 22K14029, 22H01215, and by ISelective Research Fund for Young Researchers and from Tokyo Metropolitan University.

\bibliography{main2}

\end{document}